\documentclass{IEEEtran}

\usepackage{graphicx}
\usepackage{times}
\usepackage{float}
\usepackage{eqnarray,amsmath}
\usepackage{array}
\usepackage{booktabs}
\usepackage{textcomp}
\usepackage{amsmath,amssymb,amsthm}
\usepackage{mathtools}
\usepackage{epstopdf}
\usepackage{caption}
\usepackage{multirow}
\usepackage{multicol}
\usepackage{pbox}
\usepackage{cite}
\usepackage{tikz}
\usepackage{algorithm,algpseudocode}
\usepackage{pifont}
\usepackage{dblfloatfix}
\usepackage{fixltx2e}
\usepackage{breqn}
\usepackage{subcaption}
\usepackage{xcolor}
\usepackage[font={footnotesize}]{caption}
\allowdisplaybreaks

\algnewcommand{\Inputs}[1]{%
  \State \textbf{Inputs:}\hspace*{\algorithmicindent}\parbox[t]{.8\linewidth}{\raggedright #1}
}
\algnewcommand{\Outputs}[1]{%
  \State \textbf{Outputs:}\hspace*{\algorithmicindent}\parbox[t]{.8\linewidth}{\raggedright #1}
}
\algnewcommand{\Initialize}[1]{%
  \State \textbf{Initialize:}\hspace*{\algorithmicindent}\parbox[t]{.8\linewidth}{\raggedright #1}
}
\DeclareMathOperator{\E}{\mathbb{E}}

\makeatletter
\def\endthebibliography{%
  \def\@noitemerr{\@latex@warning{Empty `thebibliography' environment}}%
  \endlist
}
\makeatother

\bstctlcite{IEEEexample:BSTcontrol}
\title{On Bounds of Spectral Efficiency of Optimally Beamformed NLOS Millimeter Wave Links}

\author{Rakesh R T, \it{Student Member, IEEE},\thanks{Copyright \textcopyright  2015 IEEE. Personal use of this material is permitted. However, permission to use this material for any other purposes must be obtained from the IEEE by sending a request to pubs-permissions@ieee.org.
		
The authors are with G. S. Sanyal School of Telecommunications,
Indian Institute of Technology, Kharagpur, W.B -India . E-mail: [rakeshrt, debarati, gdas]@gssst.iitkgp.ernet.in}~
~\normalfont{Debarati Sen}, \it{Member, IEEE}, ~\normalfont{Goutam Das} \it{}\\\vspace*{-0.1cm}}

\begin{document}
\maketitle

\begin{abstract}
Beamforming is an indispensable feature for millimeter wave (mmWave) wireless communications in order to compensate for the severe path loss incurred due to high frequency operation. In this paper, we introduce a novel framework to evaluate the spectral efficiency (SE) of non-line-of-sight (NLOS) mmWave links with optimal analog beamforming. Optimality here implies the joint selection of antenna beams at the transmitter and receiver which simultaneously maximize the received power. We develop a mathematical framework based on the extended Saleh-Valenzuela channel model to embody the impact of optimal analog beamforming into the performance metrics for NLOS mmWave links. Practical mmWave channels are characterized by sparsity in terms of number of multi-path components; we exploit this feature to derive upper and lower bounds on SE of beamformed directional links. Simulation results reveal that the proposed approach is fairly accurate to model beamformed links in most practical operating scenarios. We also study the impact of overhead due to antenna beam training on the throughput (TP) of a link and obtain an approximate solution for optimal antenna half power beamwidth which maximizes TP.
\end{abstract}


\setlength\abovedisplayskip{0pt}
\vspace{-0.25cm}
\section{Introduction}
\PARstart{R}{ecent} advances in technology have paved the way for emergence of wideband millimeter wave (mmWave) communications providing a viable option to meet the future demand for multi-Gbps data rates \cite{Rappa2}. However, high frequency mmWave transmission incurs significantly large path loss during signal propagation, and thereby limits the transmission range. To overcome this bottleneck, directional antennas with beamforming capability are employed for signal transmission and/or reception \cite{kutty2016beamforming}. The objective of beamforming protocol is to steer the antenna beams at the transmitter and receiver nodes of a link such that the transmission rate is maximized  \cite{kutty2016beamforming}. This is achieved by optimizing the signal-to-noise ratio (SNR) or signal-to-interference plus noise ratio (SINR) \cite{qiao2015mac} at the receiver.

Beamforming protocols essentially enable spatial filtering of multi-path signal components based on the defined optimality criteria \cite{qiao2015mac}. The quality and reliability of the link therefore depends on the beamformed directional channel and in this context, statistical modeling of beamformed directional channels is essential to accurately obtain mmWave network performance metrics such as coverage probability, spectral efficiency (SE) etc. The schemes proposed in \cite{bai2015coverage,di2016intensity} which evaluate the performance of mmWave networks with analog beamforming \cite{qiao2015mac} simply model the beamformed directional channel by a random gain component assuming that the channel is frequency flat. This is similar to the model used for conventional sub-6 GHz systems where channel gain is obtained as the product of a Rayleigh or Nakagami-$m$ random variable which accounts for small scale fading effect, and a path loss term that models the large scale fading effect. Similarly, a recent work on coverage analysis for mmWave line-of-sight (LOS) links with analog beamforming \cite{yu2017coverage} approximates the beamformed directional channel by a random gain component based on the uniformly random single path (UR-SP) assumption. However, in non-LOS (NLOS) mmWave channels the power content of multi-path components \cite{akdeniz2014millimeter} are comparable, and thus the modeling approaches considered for beamformed directional channels in existing literature are not applicable. Therefore, a new mathematical framework is required which embodies the impact of optimal beamforming for performance study of NLOS mmWave links.

In this paper, we develop a mathematical framework to statistically model NLOS mmWave links with optimal analog beamforming in order to evaluate the SE of noise limited NLOS mmWave links.
We assume that the optimal transmitter-receiver antenna beam pair is chosen from a set of non-overlapping antenna beams spanning the 360$^{0}$ azimuth space such that the received signal power is maximized. The omni-directional propagation characteristics of the channel is represented by the extended Saleh-Valenzuela (S-V) spatial channel model \cite{akdeniz2014millimeter,samimi201628,chen2017compressive}. We utilize this model to derive lower and upper bounds on SE of optimally beamformed mmWave links. We further note that SE of a noise limited link can be enhanced by using high resolution antenna beams albeit at the cost of significant training overhead due to the associated analog beamforming protocol \cite{shokri2015beam}. The trade-off between training overhead and throughput (TP) for indoor mmWave networks is investigated through simulations in \cite{shokri2015beam}. We propose a mathematical framework to quantify TP as a function of SE and training overhead. Moreover, the analysis also helps to determine requirements for the design of antenna beamforming protocols in terms of an optimal antenna half power beamwidth (HPBW) which maximizes TP and specifies the feasible region of operation in terms of antenna HPBW so that links are able to identify optimal antenna beam pairs. The paper has three main contributions: (i) we introduce a novel modeling approach to study the statistical behavior of optimal analog beamforming in NLOS mmWave links, (ii) we obtain tractable lower and upper bounds on SE of a NLOS mmWave link utilizing the sparsity in practical mmWave channels, and (iii) we provide a design insight for mmWave communication systems by obtaining an approximate solution for optimal antenna HPBW which maximizes TP for a given analog beamforming protocol under a set of channel parameters.
\vspace{-0.25cm}
\section{System Model}
\begin{figure}[H]\label{Deploy_scenario}
	\centering
	\includegraphics[trim=3cm 4cm 3cm 5cm, clip=true, totalheight=0.18\textheight]{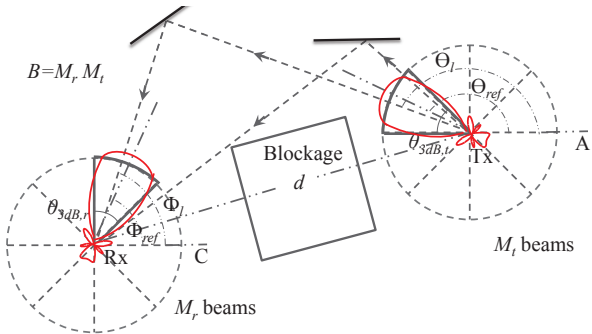}
  \caption{A typical node deployment scenario}	
	\label{fig:Deployment scenario}
\end{figure}

As shown in Fig. 1, we consider a system model consisting of an outdoor mmWave link with the transmitter and receiver nodes separated by a distance $d$. The nodes are assumed to be equipped with directional antennas with beamforming capability. We further assume that direct LOS connectivity between the transmitter and receiver is blocked and  hence the beamformed link is established through NLOS multi-path components (Fig.~\ref{fig:Deployment scenario}). We approximate the antenna radiation pattern by a sectored model \cite{bai2015coverage} with zero side lobe gain. Let $\theta_{3dB,t}$ and $\theta_{3dB,r}$ (in degrees) denote antenna HPBW of the transmitter and receiver, respectively. The transmitter and receiver main lobe antenna gain values can approximately be calculated as $G_{m,t}=\frac{360}{\theta_{3dB,t}}$ and $G_{m,r}=\frac{360}{\theta_{3dB,r}}$ \cite{bai2015coverage}, respectively. We further assume that the nodes select the optimal antenna beam pair that maximizes the SNR at the receiver node (out of $M_{t}$ and $M_{r}$ number of non-overlapping beams at the transmitter and the receiver nodes, respectively). 
\vspace{-0.35cm}
We adopt a frequency flat equivalent of the extended S-V channel model  for our analysis. The total power received from $L$ multi-path components can thus be expressed as,
\begin{align}\label{rec_pow_raw}
P=P_{T}cd^{-\alpha}\sum_{l=1}^{L}\left|h_{l}\right|^{2}G_{T}(\Theta_{ref}-\Theta_{l})G_{R}(\Phi_{ref}-\Phi_{l}),
\end{align}
where $P_{T}$ represents the transmit power, $c$ denotes the intercept point from the path loss formula, and $\alpha$ denotes the path loss exponent. $\left|h_{l}\right|$ is the small scale fading amplitude which is generally modeled as a Rayleigh or Rice random variable \cite{akdeniz2014millimeter,samimi201628}. $G_{T}(.)$ and $G_{R}(.)$ represent the antenna gain of the transmitter and receiver antennas respectively with corresponding antenna pointing angles $\Theta_{ref}$ and $\Phi_{ref}$. $\Theta_{ref}$ and $\Phi_{ref}$ are defined as the angle between the maximal gain direction of the antenna main lobe and the line segments Tx-A and Rx-C at the transmitter and receiver, respectively, as shown in Fig.~\ref{fig:Deployment scenario}. $\Theta_{l}$ and $\Phi_{l}$ are the angle of departure (AOD) and angle of arrival (AOA) of the $l$-th multi-path component. The number of multi-path components denoted by $L$ is a random variable with its average value denoted by $\lambda_{0}$ \cite{akdeniz2014millimeter}.
Assuming a sectored radiation pattern model and unit transmit power, the signal power received by the $i$-th antenna beam pair can be obtained from  (\ref{rec_pow_raw}) as $P_{i}=cd^{-\alpha}\sum_{l\in L_{i}}\left|h_{l}\right|^{2}G_{m,t}G_{m,r}$, where $L_{i}$ denotes the set of multi-path components which are located inside the antenna main lobes of the transmitter and receiver corresponding to the $i$-th antenna beam pair. We assume that the cardinality of the set $L_{i}$ ($card(L_{i})$) is a Poisson random variable with average number of multi-path components $\lambda_{d}=\lambda_{0}/B$, where $B$ denotes the total number of available transmitter-receiver beam pairs ($B=M_{t}M_{r}$). It should be noted that in practice the average received signal power varies with antenna beam orientation angle \cite{akdeniz2014millimeter}, and therefore $\lambda_{d}$ as well as $\alpha$ are functions of antenna beam orientation angle. As of now due to lack of availability of empirical data to capture this variation, we assume $\lambda_{d}$ and $\alpha$ to be constant \cite{akdeniz2014millimeter,samimi201628,chen2017compressive} which incidentally also lends mathematical tractability for analysis. It may also be noted that $\lambda_{d}$ and $\alpha$ could be obtained by making use of the analytical model reported in our prior work \cite{rakICC2017}. However, this modeling approach is presently out of scope of this paper. In this paper, the small scale fading gain $\left|h_{l}\right|$ is assumed to be Nakagami-$m$ distributed with mean power equal to $1/\lambda_{0}$, which ensures that $\E_{L,\left|h_{l}\right|^{2}}\left[\sum_{l=1}^{L}\left|h_{l}\right|^{2}\right]\approx1$, where $\E[.]$ denotes the expectation operator. The received power corresponding to the $i$-th antenna beam pair can thus be expressed as, $P_{i}=cd^{-\alpha}\lambda_{0}^{-1}\sum_{l\in L_{i}}\left|g_{l}\right|^{2}G_{m,t}G_{m,r}$ with $\E\left[\left|g_{l}\right|^{2}\right]=1$.

In practice mmWave multi-path components are sparse in time as well as the angular dimension \cite{samimi201628}. Consequently, the probability of receiving multiple propagation components inside the antenna main lobe is negligible and therefore with most of the practical antenna radiation patterns, $card(L_{i})\leq 1, \forall i$. Therefore, the presence of the multi-path component inside a pair of antenna beams can be modeled by a Bernoulli random variable with success probability $p$. The value of $p$ can be computed as, $p=1-\exp(-\lambda_{d})$. Based on this approximation, the received power corresponding to $i$-th antenna beam pair is simplified as $P_{i}=\Pi_{i}(p)\left|g\right|^{2}G_{m,t}G_{m,r}\lambda_{0}^{-1}cd^{-\alpha}$, where $\Pi_{i}(.)$ denotes the Bernoulli random variable corresponding to the $i$-th antenna beam pair with success probability $p$, i.e., $\Pi_{i}(p)=1$ with probability $p$; $\Pi_{i}(p)=0$ with probability $1-p$. The optimal transmitter and receiver antenna beams (thick lined sectors in Fig.~\ref{fig:Deployment scenario}) are jointly   selected based on the maximum received signal power criteria, and therefore the optimal received signal power is calculated as,
\begin{align}\label{sig_pow}
P_{opt}=\text{max}\left(P_{1},P_{2},...,P_{B}\right).
\end{align}
\vspace{-0.75cm}
\section{Calculation of Spectral Efficiency}
In this section, we first derive an upper bound on SE of an optimally beamformed mmWave NLOS link (included in Section III-A) by assuming Nakagami-$m$ fading for each multi-path component. In addition, we also present the upper and lower bounds on SE in Section III-A. Finally, we obtain an expression for link throughput in Section III-B which determines the fraction of SE useful for communication after accounting for the antenna beam training overhead. 
\vspace{-0.3cm}
\subsection{SE under extended S-V channel model with the assumption of Nakagami-$m$ distributed $\left|g\right|.$}
The variability in power received by each antenna beam pair is essentially due to the parameters $\left|g\right|^{2}$ and  $\Pi_{i}(p)$. Therefore, power maximization in (\ref{sig_pow}) is equivalent to the calculation of normalized received signal power corresponding to the optimal antenna beam pair, i.e., $P_{opt}^{'}=\text{max}\left(P_{1}^{'},P_{2}^{'},...,P_{B}^{'}\right),$
where  $P_{i}^{'}=\Pi_{i}(p)\left|g\right|^{2},$ $i\in\left\{1,...,B\right\}$. In this section, we first evaluate the cumulative distribution function (CDF) of $P_{opt}^{'}$. We note that $P_{i}^{'}$ for $i$-th antenna beam pair is a mixed random variable, since $\left|g\right|^{2}$ and $\Pi_{i}(p)$ are continuous and discrete random variables, respectively. Accordingly, $P_{i}^{'}$ is a continuous random variable if $\Pi_{i}(p)=1$; and a discrete random variable if $\Pi_{i}(p)=0$. This condition also implies that $P_{opt}^{'}$ is a continuous random variable if $\exists i$, where $\Pi_{i}(p)=1,\forall i\in\left\{1,...,B\right\}$. Therefore, we proceed with the derivation for the CDF of $P_{opt}^{'}$ in two exclusive parts; the first of which deals with the continuous case ($\exists i$, where $\Pi_{i}(p)=1,\forall i\in\left\{1,...,B\right\}$) and the second part deals with the discrete case ($\Pi_{i}(p)=0,\forall i\in\left\{1,...,B\right\}$) only. Hence, the CDF of $P_{opt}^{'}$ with $\exists i$, where $\Pi_{i}(p)=1,\forall i\in\left\{1,...,B\right\}$ is,\\
\begin{align}\label{prob_opt1}
F_{P'_{opt}}\left(P^{*}\right)=&Prob\left[(P_{1}^{'}\leq P^{*})\cap(P_{2}^{'}\leq P^{*})
																						...\cap(P_{B}^{'}\leq P^{*})\right.\nonumber\\&|\exists i, \text{where}\hspace{0.25cm} \Pi_{i}(p)=1,\forall i\in\left\{1,...,B\right\}\Big],
\end{align}
where $Prob(.)$ represents probability of the given event.
Since $P_{1}^{'}$,.., $P_{B}^{'}$ are independent and identically distributed, $Prob\left(P_{1}^{'}\leq P^{*}\right)$=...=$Prob\left(P_{B}^{'}\leq P^{*}\right)$=$Prob\left(P^{'}\leq P^{*}\right)$. Hence, (\ref{prob_opt1}) can be simplified based on Bayes' rule as,\\
\begin{align}\label{prob_opt3}
F_{P'_{opt}}\left(P^{*}\right)=\frac{\sum_{i=1}^{B}\binom Bi\left(1-p\right)^{B-i} p^{i} Prob(P^{'}\leq P^{*})^{i}}{1-\left(1-p\right)^{B}}
\end{align}
Without loss of generality, we calculate $Prob(P^{'}_{i}\leq P^{*})$ using the probability density function (PDF) of Gamma random variable $X$ defined as $f_{X}(x)=\frac{m^{m}x^{m-1}e^{-mx}}{\Gamma(m)}$. The CDF of $P_{i}^{'}$  $\exists i$, where $\Pi_{i}(p)=1,\forall i\in\left\{1,...,B\right\}$ is calculated as,
\vspace{0.25cm}
\begin{align}\label{rec_pow_CDF}
Prob(P_{i}^{'}\leq P^{*})=\hspace{-0.15cm}\int_{0}^{P^{*}}\hspace{-0.3cm}\frac{m^{m}x^{m-1}e^{-mx}\text{d}x}{\Gamma(m)}=\frac{\gamma(m,mP^{*})}{\Gamma(m)}
\end{align}
where $\gamma(x,y)$ denotes the lower incomplete Gamma function with parameters $x$ and $y$. Substituting $Prob(P_{i}^{'}\leq P^{*})$ in (\ref{prob_opt3}) results in,
\begin{align}\label{prob_opt4}
F_{P'_{opt}}\left(P^{*}\right)=\frac{\left(1-p\right)^{B}\left( \left[1+\frac{p}{1-p}\frac{\gamma(m,mP^{*})}{\Gamma(m)}\right]^{B}-1\right)}{1-\left(1-p\right)^{B}}
\end{align}
We note that (\ref{prob_opt4}) is intractable owing to the incomplete Gamma function. For further simplification for the computation of SE, we explore the possibility to approximate  $\frac{\gamma(m,mP^{*})}{\Gamma(m)}$ and $\left[1+\frac{p}{1-p}\frac{\gamma(m,mP^{*})}{\Gamma(m)}\right]^{B}$. Since $P^{*}$ varies from $0$ to $\infty$, only loose approximations are possible for $\left[1+\frac{p}{1-p}\frac{\gamma(m,mP^{*})}{\Gamma(m)}\right]^{B}$ which can aid the evaluation of SE.
Also, due to the possibly large values for $B$ (for example, antenna HPBWs of $33^{0}$ and $15^{0}$ at transmitter and receiver corresponds to $B=121$ and $B=625$, respectively), any approximation for the incomplete Gamma function may lead to significant error in $F_{P'_{opt}}\left(P^{*}\right)$. The only option is to minimize the error, and therefore we apply a tighter approximation, $\frac{\gamma(m,mP^{*})}{\Gamma(m)}\leq\left(1-e^{-aP^{*}}\right)^{m}$ with $a=m\Gamma(m+1)^{\frac{-1}{m}}$ \cite{bai2015coverage}. The bound on $F_{P'_{opt}}\left(P^{*}\right)$ is therefore achieved by introducing this approximation in (\ref{prob_opt4}). Further, the discrete probability component of the CDF of $P_{opt}^{'}$ is determined by the condition $\Pi_{i}(p)=0$, $\forall i\in\left\{1,...,B\right\}$. Hence, $Prob\left(\Pi^{i}(p)=0,\forall i\in\left\{1,...,B\right\}\right)=(1-p)^{B}$. The PDF of $P^{'}_{opt}$, $f_{P'_{opt}}\left(P^{*}\right)$ is obtained as,\\
\begin{align}\label{PDF_opt_pow}
f_{P'_{opt}}\left(P^{*}\right)\leq&\frac{mapB(1-p)^{B-1}}{1-(1-p)^{B}}\left(1-e^{-aP^{*}}\right)^{m-1}\nonumber\\&\times\left[1+\frac{p}{1-p}\left(1-e^{-aP^{*}}\right)^{m}\right]^{B-1}e^{-aP^{*}}
\end{align}
SE is calculated using the following formula,\\
 \begin{align}\label{SE1}
 SE=&\E\left[\text{ln}\left(1+\rho P^{'}_{opt}\right)\right]\nonumber\\=&(1-p)^{B}\text{ln}(1)+\left[1-(1-p)^{B}\right]\nonumber\\&\hspace{2cm}\times\int_{0}^{\infty}\text{ln}(1+\rho P^{*})f_{P'_{opt}}\left(P^{*}\right)\text{d}P^{*},
 \end{align}
 where $\rho=\frac{\lambda_{0}^{-1}G_{m,t}G_{m,r}cd^{-\alpha}}{\sigma^{2}}$ with $\sigma^{2}$ denotes noise power.
The upper bound on SE is obtained by substituting (\ref{PDF_opt_pow}) in (\ref{SE1}),\\
\begin{align}\label{SE2}
SE\leq&\int_{0}^{\infty}\text{ln}(1+\rho P^{*})mapB(1-p)^{B-1}\left(1-e^{-aP^{*}}\right)^{m-1}\nonumber\\&\times\left[1+\frac{p}{1-p}\left(1-e^{-aP^{*}}\right)^{m}\right]^{B-1}e^{-aP^{*}}\text{d}P^{*}.
\end{align}
The integration in (\ref{SE2}) can be evaluated by replacing $m$ with $\widehat{m}=\left\lfloor{m}\right\rfloor$. Hence, (\ref{SE2}) is modified into,\\
\begin{align}\label{SE3}
SE\leq&\sum_{i=1}^{B}(1-p)^{B}\binom Bi\left( \frac{p}{1-p}\right) ^{i}\widehat{a}\widehat{m}i\sum_{j=0}^{\widehat{m}i-1}\binom{\widehat{m}i-1}{j}\nonumber\\&\times (-1)^{j}\int_{0}^{\infty}\text{ln}(1+\rho P^{*})e^{-\widehat{a}(1+j)P^{*}}\text{d}P^{*}\nonumber\\=&\widehat{a}\widehat{m}(1-p)^{B}\sum_{i=1}^{B}\binom Bi\left( \frac{p}{1-p}\right) ^{i}i\sum_{j=0}^{\widehat{m}i-1}\binom{\widehat{m}i-1}{j}\nonumber\\&\times (-1)^{j}\frac{e^{\frac{\widehat{a}(1+j)}{\rho}}}{\widehat{a}(1+j)}E_{1}\left(\frac{\widehat{a}(1+j)}{\rho}\right),
\end{align}
where $E_{1}(.)$ denotes exponential integral function and $\widehat{a}=\widehat{m}\Gamma(\widehat{m}+1)^{\frac{-1}{\widehat{m}}}$.To provide further insights into the system design, we evaluate simplified upper and lower bounds for SE. The upper bound on SE is derived by substituting $m=1$ in (\ref{PDF_opt_pow}) (equivalent to the Rayleigh assumption for $\left|g\right|$), i.e.,\\
\begin{align}\label{PDF_Ray1}
f_{P'_{opt}}\left(P^{*}\right)=&\frac{pB}{1-(1-p)^{B}}\left(1-pe^{-P^{*}}\right)^{B-1}e^{-P^{*}}\nonumber\\\leq&\frac{pB}{1-(1-p)^{B}}\exp(-\lambda_{0}e^{-P^{*}})e^{-P^{*}}.
\end{align}
The last step in (\ref{PDF_Ray1}) is obtained from $p\approx\frac{\lambda_{0}}{B}$ (for large $B$) followed by the relation $\left(1-\frac{\lambda_{0}}{B}x\right)^{B-1}\leq e^{-\lambda_{0}x}$ \cite{abramowitz1964handbook}. Applying the inequality $\exp(-\lambda_{0}e^{-P^{*}})\leq 1-\left(1-e^{-\lambda_{0}}\right) e^{-P^{*}}$ in (\ref{PDF_Ray1}), the upper bound on SE is evaluated using (\ref{SE1}) as,\\
\begin{align}\label{SE_upper}
SE\leq pB\left[e^{\frac{1}{\rho}}E_{1}\left(\frac{1}{\rho}\right)-\frac{\left(1-e^{-\lambda_{0}}\right)}{2}e^{\frac{2}{\rho}}E_{1}\left(\frac{2}{\rho}\right)\right].
\end{align}
A closed form lower bound on SE can be derived by ignoring small scale fading for individual multi-path components. Based on the aforementioned simplification, $P^{'}_{opt}$ becomes a discrete random variable. Specifically, $P^{'}_{opt}=1$ with probability $1-\left(1-p\right)^{B}$; $0$ with probability $\left(1-p\right)^{B}$. Therefore, lower bound on SE is determined as,
\begin{align}\label{SE_lower}
SE\geq&\left(1-p\right)^{B}\text{log}(1)+\left[1-\left(1-p\right)^{B}\right]\text{ln}(1+\rho)\nonumber\\
=&\left[1-\left(1-p\right)^{B}\right]\text{ln}(1+\rho).
\end{align}
Interestingly, the bounds expressed in  (\ref{SE_upper}) and (\ref{SE_lower}) can be simplified further for highly sparse mmWave channels. Such channels are envisaged when the transmitter-receiver distance $d$ is fairly large; in fact it has been reported that the number of detectable multi-path components at the receiver decreases with transmission distance (since the power level of most multi-path components is below noise floor due to excessive propagation loss at mmWave frequencies) \cite{rappaport2015wideband}. Based on the inequality $e^{x}E_{1}(x)\leq \text{ln}(1+\frac{1}{x})$ \cite{abramowitz1964handbook} and small $\lambda_{0}$, (\ref{SE_upper}) is approximated as $SE\leq pB\text{ln}(1+\rho)\leq \lambda_{d}B\text{ln}(1+\rho)\leq\lambda_{0}\text{ln}(1+\rho).$ Similarly, the lower bound is approximated as $SE\geq\left(1-e^{-\lambda_{0}}\right)\text{ln}(1+\rho)\geq\lambda_{0}\text{ln}(1+\rho)$, which converges with the upper bound.
\vspace{-0.25cm}
\subsection{Computation of throughput by accounting antenna beam training overhead}
Generally, SE can be enhanced by operating with large $B$ which essentially increases the antenna gain. However, analog beamforming protocols require a fixed training time (with large $B$, training  time increases) to identify the antenna beam pair which maximizes SE. The training overhead reduces the opportunity of nodes to communicate due to limited residual duration for data transmission. This overhead is expected to be significant for an outdoor environment since the channel changes frequently and beamforming needs to be repeatedly performed to discover strong multi-path components. In this section, we quantify the TP of a link by associating antenna beam training overhead with SE, and derive an approximate value of optimal antenna HPBW which maximizes TP. Let $T_{o}$ denote the antenna beam training duration and let the total duration due to antenna beam training plus data transmission be $T$. In the present context, $T$ can be same as the coherence time of the channel $T_{c}$. We define TP as,\\
\begin{align}\label{TP1}
TP=\left(1-\frac{T_{o}}{T}\right)SE.
\end{align}
To evaluate TP for a practical network scenario, we consider the Multiple Sector ID Capture (MIDC)\footnote[1]{The antenna beamforming protocols may also identify sub-optimal antenna beam pairs as the solution. However, experimental evaluations confirm that the MIDC based protocol obtains the optimal antenna beam pair with fairly high probability \cite{hosoya2015multiple}.} scheme enabled antenna beamforming protocol specified in the IEEE 802.11ad standard \cite{tgad}. In the recent past, several commercial products compliant with the IEEE 802.11ad standard have been released for outdoor communications \cite{bluewireless}, although the standard was originally proposed for indoor communications.
We also note that the standard also allows nodes to employ an antenna beam tracking mechanism to track the channel variations due to mobility \cite{tgad}. The change in direction of arrival of strongest multi-path component is identified by sending a channel estimation sequence appended to the data frames. However, continuous antenna beam tracking results in reduction of data transmission duration which eventually degrades the TP of the link. Interestingly we observe that antenna beamforming (though it requires more search time compared to the antenna beam tracking mechanism) allows data transmission for a longer duration in comparison to the antenna beam tracking mechanism since it does not require any prior knowledge of the channel. As such, a judicial selection of analog beamforming and beam tracking mechanism is required for communications in highly mobile environments. However, an analysis based on this observation is presently out of scope of this paper, and we take into account analog beamforming only. 
To derive TP, we assume same number of antenna beams at transmitter and receiver nodes ($M_{t}=M_{r}=\sqrt{B}$) and $T_{o}=2\left(2\sqrt{B}+N_{b}^{2}\right)T_{f}$ \cite{hosoya2015multiple}, where $T_{f}$ represents the transmission duration of the control frame for antenna beam training and $N_{b}=4$ \cite{hosoya2015multiple}. Further, for analytical tractability, the lower bound on SE  in this section is considered with $(1-p)^{B}=\exp(-\lambda_{0})$ and $\rho=BK$, where $K=\frac{\lambda_{0}^{-1}cd^{-\alpha}}{\sigma^{2}}$. Based on this parameter setting, TP in (\ref{TP1}) is modified as,\\
\begin{align}\label{TP2}
	TP=\left(1-2\left(2\sqrt{B}+N_{b}^{2}\right)\frac{T_{f}}{T}\right)(1-e^{-\lambda_{0}})\text{ln}(1+BK).
\end{align}
The optimal value of $B$ is determined by equating the first derivative of TP from (\ref{TP2}) to zero and hence we obtain,\\
\begin{align}\label{deri_cond1}
	\frac{\left(1+\widehat{B^{*}}K\right)\text{ln}\left(1+\widehat{B^{*}}K\right)}{K\sqrt{\widehat{B^{*}}}}=\frac{1}{F_{t}}-\left(2\sqrt{\widehat{B^{*}}}+N_{b}^{2}\right),
\end{align}
where $F_{t}=\frac{2T_{f}}{T}$ and $\widehat{B^{*}}$ is the optimal value for $B$ which can be found by numerically solving (\ref{deri_cond1}). However, it is possible to derive an approximated closed form expression for $\widehat{B^{*}}$ by applying the simplification $(1+x)\text{ln}(1+x)\approx x\sqrt{x}$ in (\ref{deri_cond1}) which results in,
\begin{align}\label{deri_cond2}
	F_{t}\sqrt{K}\widehat{B^{*}}+2F_{t}\sqrt{\widehat{B^{*}}}+N_{b}^{2}F_{t}-1=0.
\end{align}
It is interesting to note that the approximation in (\ref{deri_cond1}) leads to a quadratic equation of $\sqrt{\widehat{B^{*}}}$ in (\ref{deri_cond2}). Therefore, $\sqrt{\widehat{B^{*}}}$ is,\\
\begin{align}\label{opt_num_beam}
\sqrt{\widehat{B^{*}}}=\frac{-F_{t}+\sqrt{\left(1-\sqrt{K}N_{b}^{2}\right)F_{t}^{2}+\sqrt{K}F_{t}}}{F_{t}\sqrt{K}}.
\end{align}
Correspondingly, the optimal antenna HPBW for the transmitter and receiver nodes is determined as, $\theta_{3dB,t}^{*}=\theta_{3dB,r}^{*}=\theta_{3dB}^{*}\approx\frac{360}{\sqrt{\widehat{B^{*}}}}$.
\begin{figure*}
	\centering
	\includegraphics[height=5.2cm,width=6.3cm]{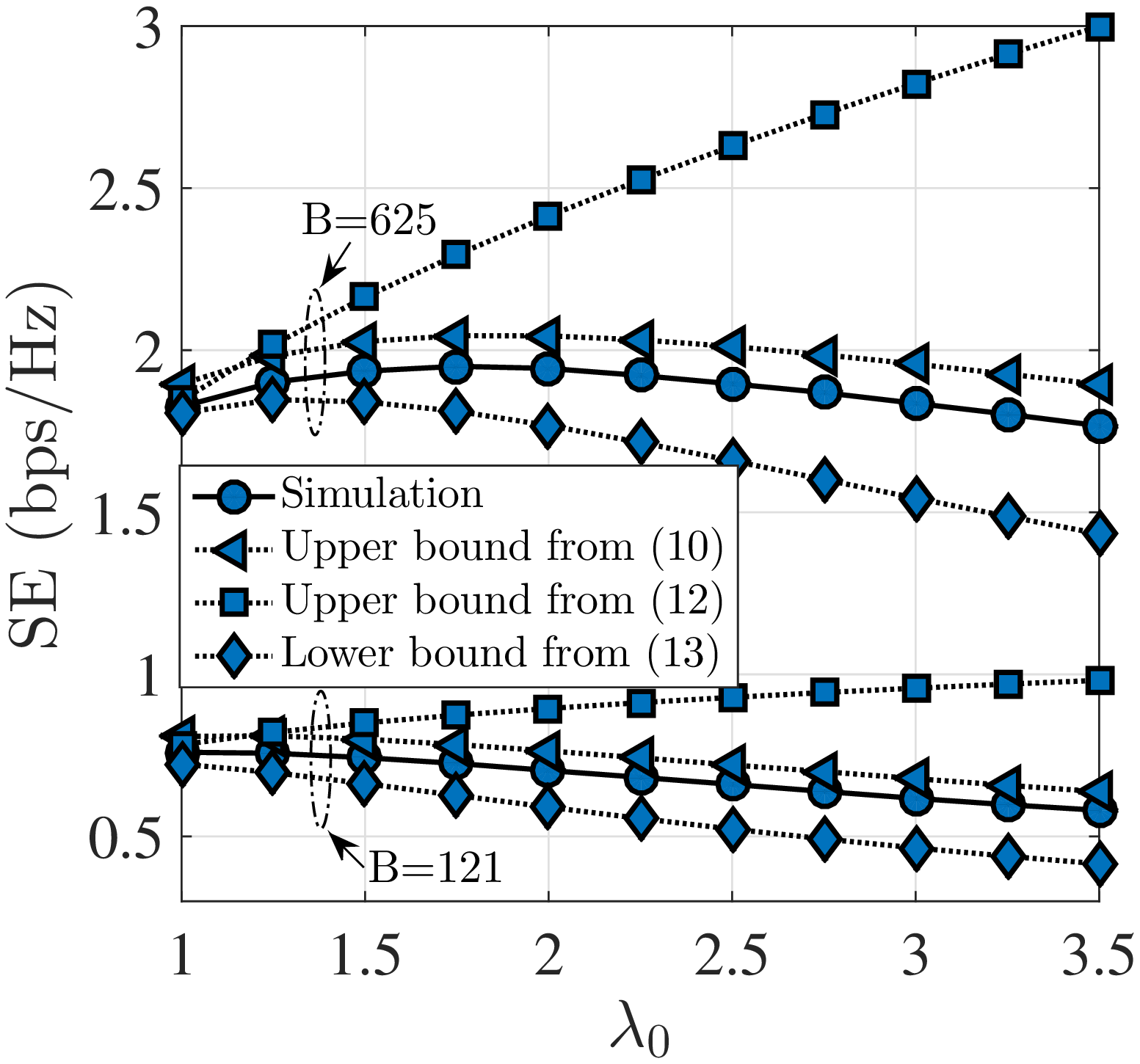}\hspace{-1.5em}
	\includegraphics[height=5.2cm,width=6.3cm]{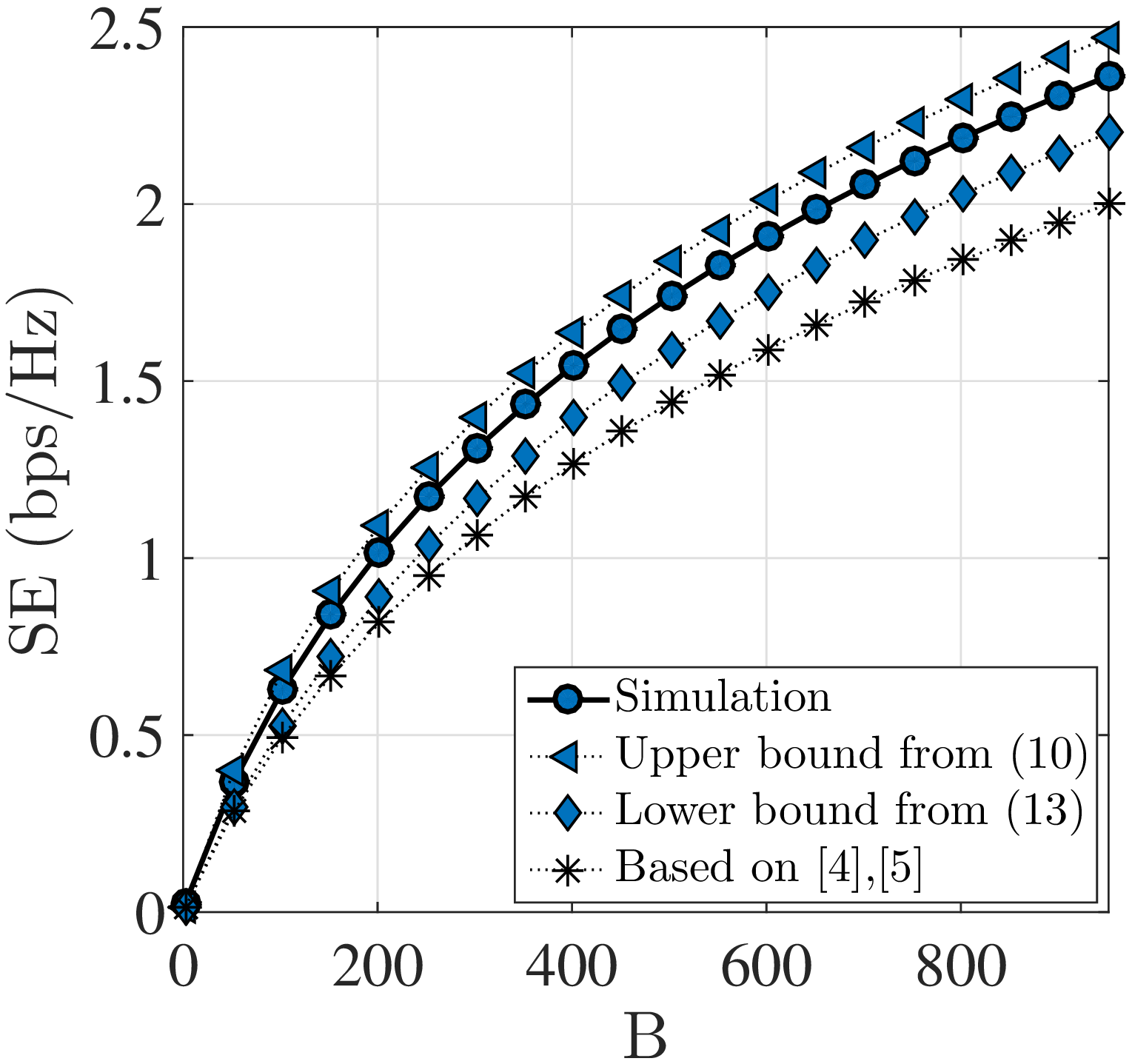}\hspace{-1.7em}	
	\includegraphics[height=5.2cm,width=6.2cm]{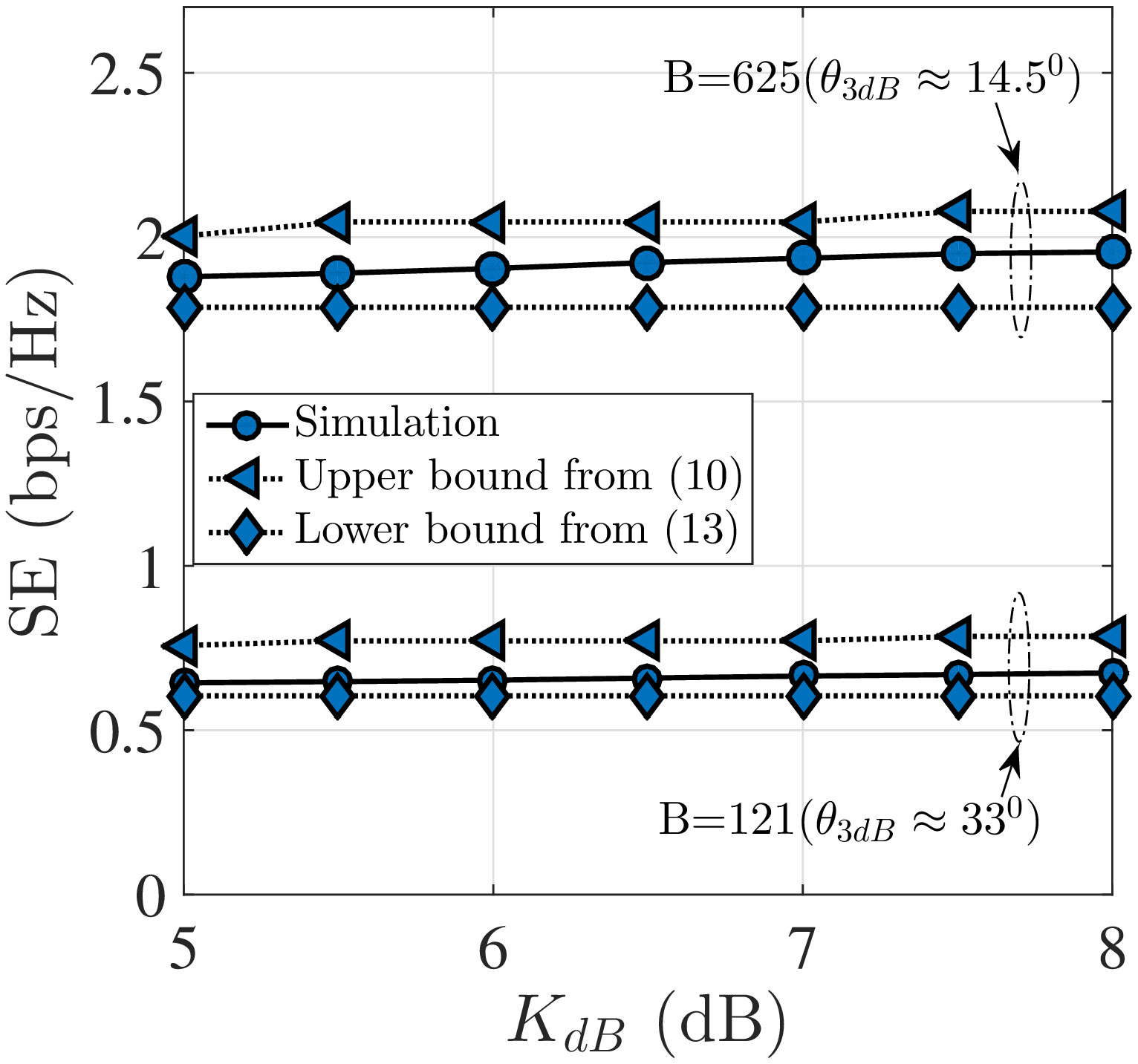}
	{$\mspace{1mu}$\small(a)$\mspace{260mu}$(b)$\mspace{300mu}$(c)}\normalsize         
	\caption{(a) Comparison of simulated plot for SE and the plots for the analytically evaluated bounds on SE for varying $\lambda_{0}$ and $m=3.2$ (b) Comparison of simulated plot for SE and the plots for the analytically evaluated bounds on SE for varying $B$, $m=3.2$ and $\lambda_{0}=1.9$ (c) Comparison of simulated plot for SE and the plots for the analytically evaluated bounds on SE for varying $K_{dB}$ with $\lambda_{0}=1.9$.}
	\label{fig:Perfor_Evaluation}
	\vspace{-0.4cm}
\end{figure*}
\section{Performance Evaluation}
Extensive Monte Carlo numerical simulations were performed to validate the analysis presented in the preceding section. We set the simulation parameters as $cd^{-\alpha}/\sigma^{2}=0.01$, $T_{f}=5$ $\mu s$ \cite{tgad}, and a variable number of antenna beam pair $B$ is selected. We consider the values of $\lambda_{0}$ spanning from 1 to 3.5 including the experimentally reported values $\lambda_{0}=1.9$ \cite{akdeniz2014millimeter} and $3.3$ \cite{rappaport2015wideband}. Firstly, we compare the simulated plot for SE and the plots for the analytically evaluated bounds on SE in Fig.~\ref{fig:Perfor_Evaluation}(a) for varying $\lambda_{o}$ and arbitrarily chosen $m=3.2$. The simulated SE plot is generated by averaging the capacity evaluated for $10^{5}$ realizations of channel. In $n$-th realization of the channel, the capacity $C_{n}$ is calculated ($C_{n}=\text{log}_{2}(1+\frac{P_{opt}}{\sigma^{2}})$) based on the criteria given in (\ref{sig_pow}), where power received in $i$-th antenna beam pair is determined as $P_{i}=cd^{-\alpha}\sum_{l\in L_{i}}\left|h_{l}\right|^{2}G_{m,t}G_{m,r}$. The variables $card(L_{i})$ and $\left|h_{l}\right|$ are generated randomly based on their respective PDFs (as discussed in the System Model section). Further, we choose two different values for $B$ in the simulation. From Fig.~\ref{fig:Perfor_Evaluation}(a), the maximum error between the upper bound derived in (\ref{SE3}) and the simulated SE is found as approximately $7.2$\% and $9.6$\% for $B=625$ and $B=121$, respectively. Moreover, the plots in Fig.~\ref{fig:Perfor_Evaluation}(a) also reveal that the upper bound from (\ref{SE_upper}) and the lower bound from (\ref{SE_lower}) are tight bounds in lower $\lambda_{0}$ regime (for example, the lower and upper bounds show error of 2.8\% and 6.1\%, respectively at $\lambda_{0}=1.25$), which indicate that the derived bounds on SE are fairly accurate for highly sparse mmWave channels.
\vspace{-0.2cm}
\begin{figure}[H]
	\centering
	\begin{subfigure}{0.24\textwidth}
		\centering
		\includegraphics[width=\textwidth]{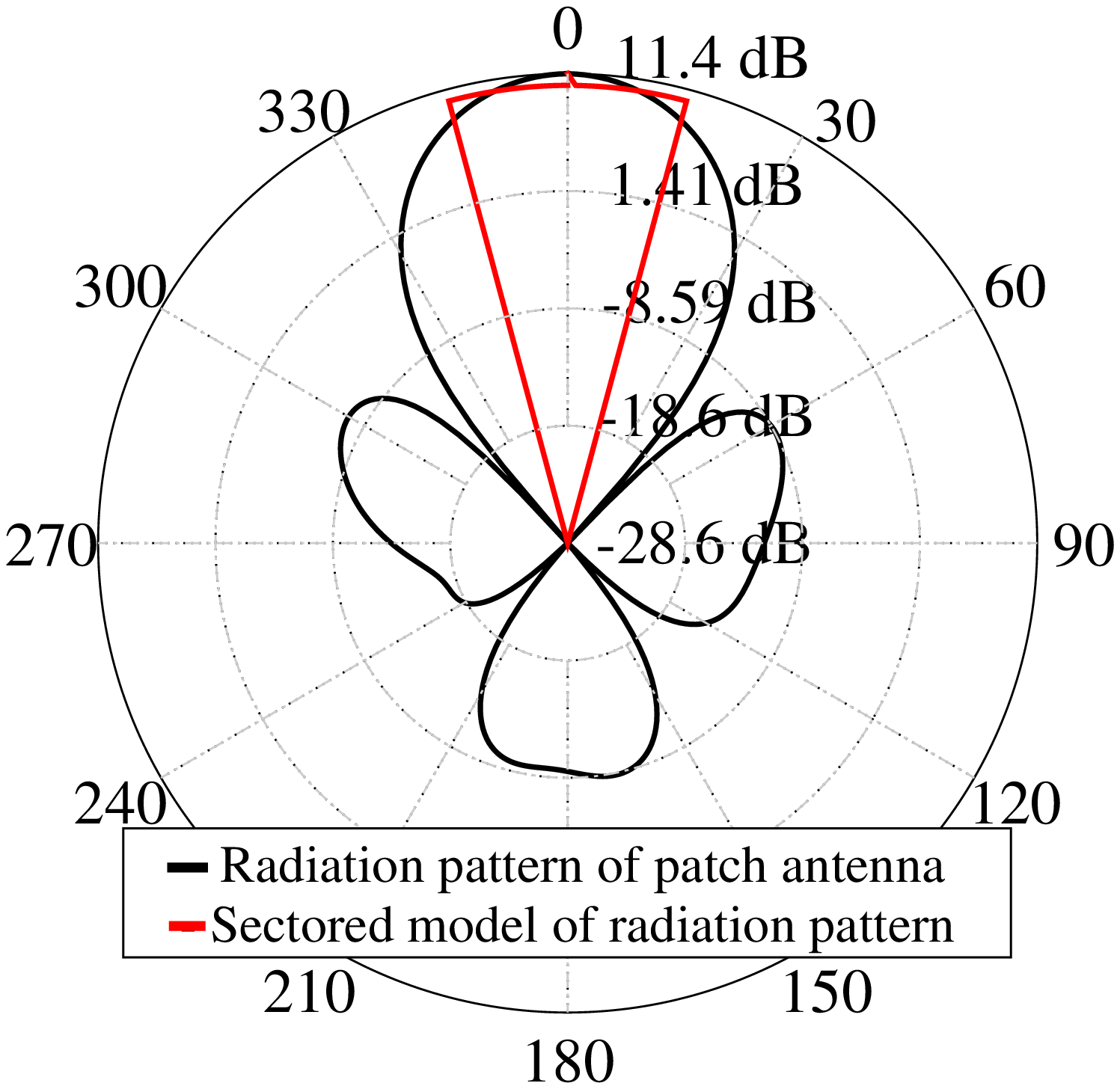}
		\vspace{-0.65cm}
		\caption{}
	\end{subfigure}
	\hspace{-0.4cm}
	\begin{subfigure}{0.24\textwidth}
		\centering
		\includegraphics[width=\textwidth]{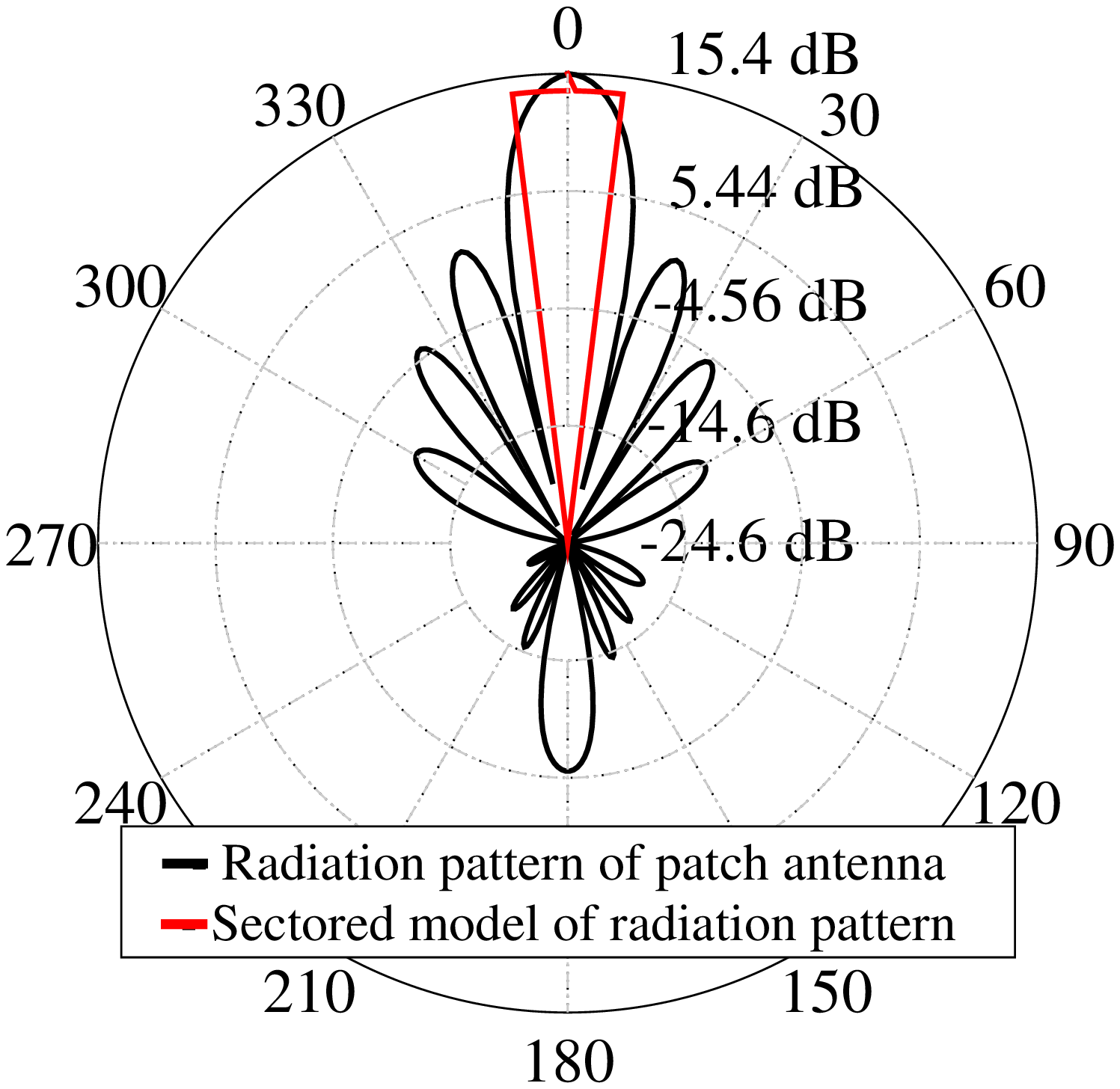}
		\vspace{-0.65cm}
		\caption{}
	\end{subfigure}%
	\caption{Radiation pattern of different patch antennas with corresponding sectored models for (a) $\theta_{3dB}=33^{0}$, (b) $\theta_{3dB}=14.5^{0}$.}
	\label{fig:exact_radiation_pattern}
\end{figure}
\vspace{-0.25cm}
Further, we compare the simulated plot for SE and the plots for the analytically evaluated bounds on SE in Fig.~\ref{fig:Perfor_Evaluation}(b) for varying $B$ with $m=3.2$ and $\lambda_{0}=1.9$. We exclude the plot for the SE bound obtained in (\ref{SE_upper}) since the corresponding error is significantly large at $\lambda_{0}=1.9$ as observed from Fig.~\ref{fig:Perfor_Evaluation}(a). For comparison, we also illustrate the result generated for the scenario where the directional channel is simply modeled as a Nakagami-$m$ ($m=2$) random variable, an assumption made in  \cite{bai2015coverage,di2016intensity}. As can be seen from Fig.~\ref{fig:Perfor_Evaluation}(b), the error between upper bound in (\ref{SE3}) and the simulated SE is reduced from approximately $8.7$\% to $4.6$\% for $B=100$ ($\theta_{3dB}=\theta_{3dB,t}=\theta_{3dB,r}=36^{0}$) and $B=1000$ ($\theta_{3dB}=\theta_{3dB,t}=\theta_{3dB,r}=11.4^{0}$). We also observe that the error due to the lower bound from (\ref{SE_lower}) is as large as 15.6\% for $B=100$. However, for higher values of $B$, the error is seen to reduce to 6.8\%, which indicates that the closed form lower bound may also applicable
for SE evaluation of optimally beamformed links for higher values of $B$ or equivalently while operating with low resolution antenna beams.
\vspace{-0.25cm}
\begin{figure}[H]
	\centering
	\includegraphics[height=5.8cm,width=8.2cm]{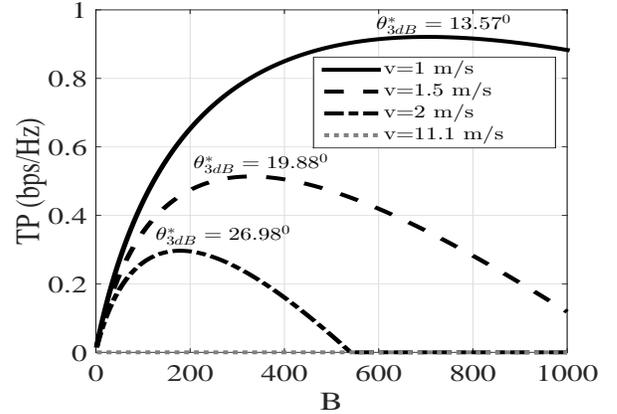}
	\caption{Effect of channel dynamics on TP}
	\label{fig:optimum_antenna_beamwidth}
\end{figure}
\vspace{-0.25cm}
The plot for simulated SE assuming Rician distributed $\left|h_{l}\right|$ is presented in Fig.~\ref{fig:Perfor_Evaluation}(c). In the same figure we also plot the bounds on SE (analytical) for varying Rician shape parameter, $K_{dB}$ and $\lambda_{0}=1.9$. Experimentally reported values of $K_{dB}$ \cite{samimi201628} were used in the simulation. The analytical plot corresponding to the upper bound in (\ref{SE3}) is generated for the values of $m$ determined from chosen $K_{dB}$ using (2.54) in \cite{stuber2001principles}. An omni-directional channel is generated in each iteration of the simulation with Rician distributed multi-path amplitude for adopted shape parameter $K_{dB}$. Antenna radiation pattern is then applied (based on discrete set of antenna pointing directions) to obtain optimally beamformed directional channel. Two different patch antenna radiation patterns with antenna HPBW, $\theta_{3dB}=\theta_{3dB,t}=\theta_{3dB,r}=14.5^{0}$ and $33^{0}$ depicted in Fig.~\ref{fig:exact_radiation_pattern} are used for simulation. Correspondingly, we choose $B=625$ and $B=121$ for the analytical plots assuming that each node employs the same antenna HPBW ($14.4^{0}$ and $32.73^{0}$, respectively) for communication. As shown in Fig.~\ref{fig:Perfor_Evaluation}(c), a maximum error of $8.2$\% and $18.5$\% is observed for $B=625$ and $B=121$, respectively as a result of the joint impact of the simplification of antenna radiation pattern with sectored model and the approximations adopted for the evaluation of the bound. We also note that the maximum error between the simulated SE and the lower bound on SE is only 10.2\% for $B=121$.

Finally in Fig.~\ref{fig:optimum_antenna_beamwidth}, we plot simulated TP of a link with SE determined using the simulation procedure used for Fig.~\ref{fig:Perfor_Evaluation}(a) and Fig.~\ref{fig:Perfor_Evaluation}(b). We use $T=T_{c}$ for the simulation.
To the best of our knowledge, an investigation pertaining
to channel dynamics in outdoor millimeter wave environment is
unavailable in literature. Therefore, $T_{c}$ is analytically determined by assuming that the channel dynamics is only due to the motion of transmitter or receiver node alone (related mathematical expressions are available in \cite{smulders2009}). In Fig.~\ref{fig:optimum_antenna_beamwidth}, we plot TP for four scenarios. For three scenarios, one of the nodes is assumed to be carried by a moving person and the in the last scenario, the node is assumed to be located in a moving vehicle (accordingly, velocity of the node varies from $v=1$ m/s to $v=11.1$ m/s ($=$ $3.6$ km/hr $-$ $40$ km/hr)). As shown in Fig.~\ref{fig:optimum_antenna_beamwidth}, TP initially increases and then reduces owing to the increasing training overhead due to antenna beamforming. Moreover, when velocity of the node increases, the maximum achievable TP reduces. As evident from Fig.~\ref{fig:optimum_antenna_beamwidth}, there also exists a range of $B$ for which TP of the link becomes negative (corresponding values of TP are truncated to zero in Fig.~\ref{fig:optimum_antenna_beamwidth}). We note that it is not possible to complete the beamforming procedure  within the duration of $T_{c}$, a fact also evident from (\ref{TP2}). Consequently, for the nodes employing fixed antenna radiation pattern, beamforming protocol may have to choose a sub-optimal antenna beam pairs so that those nodes are able to commence data transmission before the channel starts changing significantly. This effect is severe for highly mobile environments as shown by the plot corresponding to $v=11.1$ m/s, which reveals that the identification of optimal antenna beam pairs is not possible no matter what is the value of $B$. We also calculate approximated $\theta_{3dB}^{*}$ using the formula $\theta_{3dB}^{*}\approx \frac{360}{\sqrt{\widehat{B^{*}}}}$. The calculated values are $\theta_{3dB}^{*}\approx 13.16^{0}, 18.32^{0},$ and $23.93^{0}$, respectively for $v=1,1.5,$ and $2$ m/s and the corresponding  simulated values are $\theta_{3dB}^{*}=13.57^{0}, 19.88^{0},$ and $26.98^{0}$. Thus, the analytical framework presented in Section III-B can serve as a design tool for beamforming protocols for outdoor mmWave communications.
\vspace{-0.25cm}
\section{Conclusion}
In this paper, we present a novel methodology to incorporate optimal analog beamforming into the framework for evaluation of bounds on SE of NLOS mmWave links. We establish that the simplistic assumption of Rayleigh or Nakagami-$m$ probability distribution for the beamformed directional channel gain is inadequate to characterize NLOS mmWave beamformed channels. In addition, we also investigate the effect of antenna beam training overhead on throughput of a link, and identify the necessary conditions for its maximization. The evaluation of throughput based on an standard antenna beamforming protocol reveals that for a highly mobile environment, it may not be possible to identify optimal antenna beam pairs which maximize SE, and as such the nodes may end up operating with sub-optimal antenna beams.

As future work, it would be interesting to explore the scenario where average number multi-path components per antenna beam and the path loss exponent is considered to be a function of the antenna beam orientation angle.
\vspace{-0.25cm}
\setcounter{secnumdepth}{0} 
\section{Acknowledgment} 
The author would like to thank to Mr. Shajahan Kutty for proof reading this manuscript, and also express gratitude to Dr. Saswati Ghosh for providing simulated antenna radiation patterns for analysis.


\begin{thebibliography}{9}
	\bibitem{Rappa2}
	T. S. Rappaport, R. W. Heath Jr, R. C. Daniels, and J. N. Murdock, \textit{Millimeter Wave Wireless Communications}. Prentice Hall Communications Engineering and Emerging Technologies Series, 2014.
	
	\bibitem{kutty2016beamforming}
	S. Kutty and D. Sen, ``Beamforming for millimeter wave communications: An inclusive survey'', \textit{IEEE Commun. Surveys Tuts.}, vol. 18, no. 2, pp. 949-973, 2016.
	
	\bibitem{qiao2015mac}
	J. Qiao, X. Shen, J. W. Mark, and Y. He, ``MAC-layer concurrent beamforming protocol for indoor millimeter-wave networks'', \textit{IEEE Trans. Veh. Technol.}, vol. 64, no. 1, pp. 327-338, 2015.
	
	\bibitem{bai2015coverage}
	T. Bai and R. W. Heath, ``Coverage and rate analysis for millimeter-wave cellular networks'', \textit{IEEE Trans. Wireless Commun.}, vol. 14, no. 2, pp. 1100-1114, 2015.
	
	\bibitem{di2016intensity}
	M. Di Renzo, W. Lu, and P. Guan, ``The intensity matching approach: A tractable stochastic geometry approximation to system-level analysis
	of cellular networks'', \textit{IEEE Trans. Wireless Commun.}, vol. 15, no. 9, pp. 5963-5983, 2016.
	
	\bibitem{yu2017coverage}
	X. Yu and J. Zhang \textit{et al.}, ``Coverage Analysis for Millimeter Wave Networks: The Impact of Directional Antenna Arrays'',
	\textit{IEEE J. Select. Areas Commun.}, vol. 35, no. 7, 2017.
	
	\bibitem{akdeniz2014millimeter}
	M. R. Akdeniz, and Y. Liu \textit{et al.}, ``Millimeter wave channel modeling and cellular capacity evaluation'', \textit{IEEE J. Select. Areas Commun.}, vol. 32, no. 6, pp. 1164-1179, 2014.
	
	\bibitem{samimi201628}
	M. K. Samimi and G. R. MacCartney \textit{et al.}, ``28 GHz millimeter-wave ultrawideband small-scale fading models in wireless channels'', in \textit{IEEE 83rd Vehicular Technology Conference (VTC Spring)},	2016, pp. 1-6.
	
	\bibitem{chen2017compressive}
	C.-H. Chen and C.-R. Tsai \textit{et al.}, ``Compressive Sensing (CS) Assisted Low-Complexity Beamspace Hybrid Precoding for Millimeter-Wave MIMO Systems'', \textit{IEEE Trans. Signal Process.}, vol. 65, no. 6, pp. 1412-1424, 2017.
	
	\bibitem{shokri2015beam}
	H. Shokri-Ghadikolaei, L. Gkatzikis, and C. Fischione, ``Beam-searching	and transmission scheduling in millimeter wave communications,'' \textit{in
	Proc. IEEE International Conf. on Comm.}, 2015, pp. 1292-1297.
	
	\bibitem{rakICC2017}
	R. T. Rakesh, G. Das, and D. Sen, ``An Analytical Model for Millimeter Wave Outdoor Directional Non-Line-of-Sight Channels'', \textit{in Proc. IEEE International Conf. on Comm.}, 2017, pp. 1-6.
	
	\bibitem{abramowitz1964handbook}
	M. Abramowitz and I. A. Stegun, \textit{Handbook of mathematical functions:	with formulas, graphs, and mathematical tables.} Courier Corporation, 1964.
	
	\bibitem{rappaport2015wideband}
	T. S. Rappaport and G. R. MacCartney \textit{et al.}, ``Wideband millimeterwave propagation measurements and channel models for future wireless communication system design,''  \textit{IEEE Trans. Commun.,} vol. 63, no. 9, pp. 3029-3056, 2015.
	
	\bibitem{hosoya2015multiple}
	K. Hosoya and N. Prasad \textit{et al.}, ``Multiple sector ID capture (MIDC): A novel beamforming technique for 60-GHz band multi-Gbps WLAN/PAN systems,” \textit{IEEE Trans. Antennas Propag.}, vol. 63, no. 1, pp. 81-96, 2015.
	
	\bibitem{tgad}
	``Part 11: Wireless LAN Medium Access Control (MAC) and Physical Layer (PHY) Specifications Amendment 3: Enhancements for Very High Throughput in the 60 GHz Band,'' December 2012.
	
	\bibitem{bluewireless}
	http://www.bluwirelesstechnology.com/product/.
	
	\bibitem{stuber2001principles}
	G. L. St{\"u}ber, \textit{Principles of mobile communication.} Springer, 2001.
	
	\bibitem{smulders2009}
	P. F. Smulders, ``Statistical characterization of 60-GHz indoor radio channels,'' \textit{IEEE Trans. Antennas Propag.,} vol. 57, no. 10, pp. 2820-2829, 2009.
\end{thebibliography}
\end{document}